# Pressure-Tuned Competing Electronic States in Layered Tellurides

Mahmoud Abdel-Hafiez[1,2], Govindaraj Lingannan[2], D. A. Chareev[3,4], A. N. Vasiliev[4], Anas Abutaha[5], Kadir Can Dogan[6], Mehmet Yagmurcukardes[7], Mehmet Egilmez[8,9], Hasan Sahin[8,9], Sami El-Khatib[8,9]

[1]Department of Applied Physics and Astronomy, University of Sharjah, PO Box 27272, Sharjah, UAE.
[2]College of General Education, University of Doha for Science and Technology, Doha, Qatar.
[3]Korzhinskii Institute of Experimental Mineralogy RAS, Chernogolovka 142432, Russia.
[4]Ural Federal University, Ekaterinburg 620002, Russia.
[5]Qatar Environment and Energy Research Institute (QEERI), Hamad Bin Khalifa University, Doha 34110, Qatar.
[6]Department of Physics, Gulbahce Campus, Izmir Institute of Technology, Izmir, Turkey.
[7]Department of Photonics, Gulbahce Campus, Izmir Institute of Technology, Izmir, Turkey.
[8]Department of Physics, American University of Sharjah, PO Box 26666, Sharjah, UAE.
[9]Materials Research Center, American University of Sharjah, PO Box 26666, Sharjah, UAE

Layered transition-metal dichalcogenides (TMDs) host competing electronic states that can be tuned by external perturbations, providing a platform to explore the interplay between disorder, electronic structure, and quantum transport. Here we investigate magnetotransport in bulk semiconducting 2H-MoTe$_2$ under hydrostatic pressure. At ambient pressure, transport evolves from high-temperature metallic behavior into activated conduction and ultimately a strongly localized variable-range hopping regime, accompanied by a pronounced magnetotransport anomaly near 45 K and large, nonsaturating magnetoresistance extending up to an unprecedented field of 60 T in semiconducting 2H-MoTe$_2$. Under compression to 15.6 GPa, the insulating state is rapidly suppressed and a low-resistivity regime emerges in which quantum interference dominates, exhibiting a crossover from weak antilocalization (WAL) to weak localization (WL) at low temperatures. A physically motivated phenomenological description captures the magnetoresistance across these regimes and yields a characteristic electronic length scale that remains comparable across the localized and quantum-interference regimes. First-principles calculations reveal a continuous pressure-driven collapse of the bandgap into a semimetallic electronic structure. These results establish a unified picture of pressure-tuned transport spanning hopping and quantum-coherent regimes.

Molybdenum ditelluride ($MoTe_2$) is a TMD hosting multiple crystallographic polymorphs that lie unusually close in energy [1-4]. In the semiconducting 2H phase, Mo atoms are trigonal-prismatically coordinated within strongly covalent Te–Mo–Te layers, while adjacent layers are weakly coupled through van der Waals interactions (inset to Fig. 1(b)). In contrast, distorted octahedral coordination stabilizes the 1T′ structure, which transforms into the orthorhombic Td phase with broken inversion symmetry. Beyond the 2H phase, $MoTe_2$ hosts metallic, semimetallic, and topologically nontrivial electronic states [1-6].

Existing work has primarily focused on chemically unstable 1T′-$MoTe_2$ films [7] and defect-driven transport in 2H-$MoTe_2$ associated with Te deficiency [8], while related investigations in $MoSe_2$ have revealed hopping transport and negative magnetoresistance [9,10]. In parallel, Weyl semimetallic and quantum spin Hall states, together with constructive/destructive quantum-interference transport, have been reported in 1T′-$MoTe_2$ [11,12], with further studies demonstrating magnetotransport crossover in $WS_2$ and thickness-dependent transport in 1T′ and Td $MoTe_2$ [13-15]. Transport in $MoTe_2$ is further shown to be highly sensitive to thermomechanical, electrostatic, and charge-mediated tuning [16-18], while large, often non-saturating magnetoresistance is characteristic of semimetallic $MoTe_2$ and $WTe_2$ systems [19-24]. External perturbations such as pressure, strain, and chemical modification can strongly modify the electronic structure in $MoTe_2$ systems [25-27], yet a unified transport-level description within a single 2H-$MoTe_2$ crystal remains lacking. Here we address this gap by combining electrical transport measurements with first-principles calculations to track the evolution of transport in bulk 2H-$MoTe_2$ under hydrostatic pressure.

High-quality bulk 2H-$MoTe_2$ single crystals were grown by chemical vapor transport following established protocols [28,29]. Single-crystal and out-of-plane X-ray diffraction were performed using a Bruker D8 Venture (Mo radiation) and a Rigaku SmartLab (Cu K$_\alpha$), respectively, to resolve (00$\ell$) reflections and confirm *c*-axis orientation. Single-crystal XRD yields lattice parameters a = b = 3.5206(2) Å, c = 13.9667(12) Å, with unit-cell volume 149.919(18) Å$^3$ and angles α = β = 90°, γ = 120°, consistent with prior reports [30]. Raman measurements were carried out on a Renishaw inVia spectrometer (514 nm excitation). Electrical transport measurements were conducted in a Quantum Design PPMS over 1.8–400 K with magnetic fields up to 9 T. High-field magnetotransport up to 60 T was performed at a pulsed-field facility using standard acquisition techniques [29,31,32]. Crystal orientation was verified by Laue diffraction. High-pressure transport was carried out in a nonmagnetic diamond anvil cell using a four-probe configuration with Pt leads. Samples (∼70×70×10 μm$^3$) were loaded into a cubic boron nitride/epoxy insulating gasket with Daphne 7373 as the pressure medium. Pressure was calibrated via ruby fluorescence. Pressure-dependent transport measurements were performed in the PPMS.

First-principles calculations were performed within density functional theory using the Vienna *ab initio* Simulation Package (VASP) [33]. The interaction between valence electrons and ionic cores was treated using the projector-augmented wave method [34], and exchange–correlation effects were described within the generalized gradient approximation using the Perdew–Burke–Ernzerhof functional [35]. Spin orbit coupling was explicitly included in all electronic structure calculations. Long-range dispersion interactions were accounted for using the semi-empirical DFT-D2 correction [36]. A plane-wave kinetic energy cutoff of 500 eV was employed. Structural relaxations were carried out using a Γ-centered 10×10×1 Monkhorst–Pack *k*-point mesh [37], with total energies converged to 10$^{-5}$ eV and residual Hellmann–Feynman forces reduced below 10$^{-4}$ eV Å$^{-1}$.

Figs. 1(a) and 1(b) show the XRD pattern measured along the crystallographic *c*-axis and Raman response, confirming the layered structure and high crystalline quality of bulk 2H-MoTe$_2$ [1-3,30]. The out-of-plane ambient pressure temperature dependence of the resistivity ($\rho_c(T)$) is presented in Fig. 1(c). Upon cooling from 400 K to 1.8 K, $\rho_c(T)$ exhibits semiconducting behavior, with a weak upward curvature emerging near 130 K [1-3,5,25,27,38]. $\rho_c(T)$ in the $\frac{d\rho}{dT} > 0$ region is well described by a dominant Fermi liquid (FL) *e–e* $T^2$ term, supplemented by $T$ and $T^3$ *e–ph* scattering contributions, and a high-fidelity fit labeled FL is shown as a solid red line. At low temperatures (6-30 K), $\rho_c(T)$ signals strong carrier localization and is described by a variable-range-hopping (VRH) captured by $\rho(T) = \rho_0 e^{\left(\frac{T_0}{T}\right)^p}$, where $\rho_0$, $T_0$, and $p$ are fitting parameters. The solid (blue, labeled VRH) line returns a 3D Mott VRH with $p = 0.245$, and a characteristic temperature $T_0 = 2.2 \times 10^6$ K [1,4,5,9,10,39]. Extrapolating the independently extracted high and low temperature transport parameters across the full 1.8–400 K range yields a consistent description of charge transport in bulk 2H-MoTe$_2$, which can be summarized as follows: (i) A simple series combination of the metallic and strongly localized contributions reproduces the entire experimental $\rho_c(T)$ (inset of Fig. 1(c)) [40-42]. (ii) The resistivity in the intermediate temperature range (30–120 K) is consistent with a percolative transport regime, as it follows an Arrhenius form with an activation energy $\Delta \sim 10$ meV, in which locally conducting regions or impurity-band states begin to connect. (iii) A large $T_0$ ($\sim 10^6$ K) implies a short localization length and a disorder-dominated energy scale, yielding an effective transport gap of ~10.6 meV at 30 K. The consistency of these energy scales indicates that the apparent activation energy emerges from the same disorder-defined landscape governing hopping transport in the low-temperature regime [39,43,44]. (iv) Finally, extrapolation of the dominant scattering mechanisms beyond their fitting ranges (red and blue dashed lines in Fig. 1(c)) reveals a crossing near 45 K. In bulk 2H-MoTe$_2$, this crossover is consistent with the onset of disorder-dominated, percolative conduction, potentially enhanced by stacking faults, local structural distortions, or incipient polytypism inherent to weakly coupled van der Waals layers [45,46].

Fig. 1(d) shows a clear magnetoresistive response ρ($T$) under 0 and 9 T applied parallel to the $c$-axis (black open/solid circles) and within the $ab$-plane (blue open/solid squares) over the full temperature range. These measurements extend beyond previously explored temperatures (e.g., Ref. [5]). ρ($T$) shows weak anisotropy, consistent with the quasi-layered nature of MoTe$_2$ [21]. The right axis magnetoresistance ($MR(\%) = \frac{\rho(H)-\rho(0)}{\rho(0)} \times 100\%$) in Fig. 1(d) exhibits a maximum near 45 K, coincident with the crossing temperature extracted in Fig. 1(c). This concurrence identifies the intermediate temperature regime as the point of maximal sensitivity of charge transport, where the applied field perturbs carrier trajectories and connectivity, consistent with a competition between extended and localized transport pathways, rather than a simple orbital or geometric magnetoresistance [47,48]. Although bulk 2H-MoTe$_2$ is nonmagnetic in its pristine form, intrinsic point defects, including tellurium vacancies, metal vacancies, and chalcogen-metal antisites can give rise to magnetic-like responses below ~50 K, as evidenced by muon spin rotation and susceptibility measurements [26,27]. We propose that such local defects disrupt covalent bonding and generate spatially inhomogeneous electronic environments, providing a plausible microscopic origin for enhanced field sensitivity.

The MR($H$) remains nonsaturating up to ~60 T in Fig. 2(a), extending beyond previously reported field ranges and reinforcing the prominence of the 45 K scale. It exhibits a pronounced anisotropic response within the intermediate temperature range, where MR reaches 800% at 45 K for $H//ab$-plane, while a reduced but qualitatively similar response is observed for $H//c$-axis. Nonsaturating magnetoresistance has been widely reported in layered semimetals and TMDs, often in conjunction with extremely large MR and quantum oscillations [2,13,16,20-22,24,26]. In compensated bismuth and graphite semimetals, large MR can be attributed to a two-band framework, and saturation is expected at sufficiently high magnetic fields due to deviations from perfect compensation and the approach to the quantum limit [49-51]. In contrast, materials such as $WTe_2$ exhibit unusually robust nonsaturating MR as a consequence of nearly perfect electron-hole compensation over an extended field range [19]. Our results place bulk $2H-MoTe_2$ in a distinct transport regime, where nonsaturating MR extends to 60 T in a narrow-gap semiconducting regime, driven by electronic inhomogeneity and multichannel transport rather than intrinsic semimetallicity.

Deep within the 3D Mott VRH regime, the MR exhibits features that have not been previously explored in semiconducting $2H-MoTe_2$. MR($T$) at 9 T in the range of 13–400 K is summarized in Fig. 2(b), revealing a discernible anisotropy, reaching a maximum near 45 K, and gradually diminishing on both sides. Representative MR($H//c$) curves are presented in Figs. 2(c)-(d). Above 18 K, Fig. 2(c) shows a positive MR in both the metallic and insulating regimes, supporting the 45 K feature. In the metallic regime, the positive parabolic MR arises, where the Lorentz force bends carrier trajectories and increases the resistivity in this nonmagnetic system [39]. Upon cooling below the crossover near 45 K, where the resistivity rises rapidly and transport becomes insulating, the magnetoresistance remains positive, consistent with an incoherent transport regime in which band-like motion is suppressed and quantum interference corrections have not yet developed [41]. The persistence of a positive MR response on both sides of the crossover reflects the fact that distinct scattering mechanisms can produce the same MR sign [39,41]. Cooling below 18 K, the MR in Fig. 2(d) develops a clear negative contribution at low fields. The negative component progressively strengthens below 15 K and eventually overwhelms the positive contribution. Such behavior cannot be accounted for within a diffusive WL and/or WAL framework, reported for metallic 1T′ MoTe$_2$ [11-13,15,23], and is inconsistent with the strong 3D VRH localization discussed above. Instead, the evolution of the low field MR reflects a crossover from conventional orbital contributions at elevated temperatures to interference-controlled hopping transport at low temperatures. The MR reflects the competition between quantum-interference effects intrinsic to VRH and conventional field-induced suppression of hopping transport. At low fields, quantum interference between alternative hopping paths gives rise to a negative MR that is quadratic in field, as described by the Nguyen-Spivak-Shklovskii interference model and further developed by Sivan-Imry-Entin-Wohlman [52-54]. As the field increases, this negative contribution progressively saturates once the magnetic flux threading a typical hopping loop becomes comparable to a flux quantum, indicating the suppression of interference effects. At higher fields, a positive quadratic MR emerges, originating from magnetic-field-induced shrinkage of localized electronic wavefunctions, consistent with the Efros-Shklovskii (ES) framework for hopping transport [41]. To capture this crossover behavior, the MR is analyzed using the phenomenological form:

$$MR(\%) = -A \frac{H^2}{1 + \frac{H^2}{H_{\text{loop}}^2}} + PH^2 \quad (1)$$

The first term describes the quantum negative MR that saturates above a characteristic crossover field $H_{\text{loop}}$, while the second term accounts for the positive wavefunction shrinkage. $A$, $P$, and $H_{\text{loop}}$ are fitting parameters. $H_{\text{loop}}$ defines the field scale at which quantum interference between alternative hopping paths is effectively destroyed and is related to an effective loop size $L_{\text{loop}} \sim \sqrt{\phi_0/H_{loop}}$, where $\phi_0$ is the magnetic flux quantum ($= h/e$, where $h$ is Planck's constant, and $e$ is the elementary charge). The fits to Eq. (1) are shown by solid red lines in Fig. 2(d), and the extracted fitting parameters $P$, $A$, and $L_{\text{loop}}$ in Figs. 2(e)-(g) reflect the field-induced hopping suppression with increasing $T$. Within the ES-VRH, a localization length $\xi$, in the hopping window can be extracted when the positive MR is observed, as expected from wave-function shrinkage [41,54] where:

$$\ln\left(\frac{R(H)}{R(0)}\right) = C_M \left(\frac{eH}{\hbar}\right)^2 \xi^4 \left(\frac{T_0}{T}\right)^{\frac{3}{4}} \qquad (2)$$

where $C_M$ is a dimensionless fitting coefficient, $T_0$ is the characteristic Mott temperature for 3D VRH, $\xi$ is the localization length, $e$ is the elementary charge, and $\hbar$ is the reduced Planck's constant. The fit at $T = 25$ K is shown as a solid red line in the inset of Fig. 2(c), and the extracted $\xi \approx 1.99$ nm. The combination of $T_0$ and $\xi$ provides an estimate of the average hopping length through $r_h = C_R \xi \left(\frac{T_0}{T}\right)^{\frac{1}{4}}$ [54,55], where $C_R \simeq 0.5$–$1$ for 3D Mott VRH, and the extracted $r_h \sim 15$–$30$ nm, consistent with hopping within the characteristic size $L_{\text{loop}}$. Our extracted $\xi \sim 1.99$ nm, $r_h \sim 30$ nm, and $L_{\text{loop}} \sim 57$ nm in 2H-MoTe$_2$ form a consistent set of characteristic scales for disorder-controlled hopping transport: charge carriers remain confined on the scale of $\xi$, undergo thermally activated hops over distances $r_h$, and explore closed interference paths of extent $L_{\text{loop}}$ before phase coherence is lost. This indicates a disorder-fragmented electronic landscape in which transport proceeds via multi-step hopping among defect-defined clusters, producing magnetoresistance governed by interference among alternative tunneling trajectories. Small localization lengths (~1.8 nm) and MR crossovers driven by disorder have been reported in bilayer MoSe$_2$ and multilayer WS$_2$ [10,14]. Moreover, MoTe$_2$ films exhibit large $T_0$ and small $\xi$ arising from grain boundaries and chalcogen deficiency, confirming the crucial role of intrinsic disorder in 2H-TMDs [1].

Hydrostatic pressure on bulk 2H-MoTe$_2$ has a dramatic impact on $\rho_c(T)$ as shown in Fig. 3(a). The insulating behavior (at 0 GPa) is rapidly suppressed, and the transport curves evolve continuously toward metallic-like behavior at the highest measured pressure of 15.6 GPa, in line with previous studies [25,56,57]. This pronounced pressure sensitivity highlights the fragile nature of the ambient-pressure semiconducting state and underscores the role of lattice compression in reshaping the electronic transport in a manner consistent with proximity to a semimetallic electronic manifold associated with topologically nontrivial phases. $\rho_c(T)$ at 15.6 GPa reveals distinctive features as shown separately in Fig. 3(b). A clear anomaly appears at 225 K in $\rho_c(T)$, resembling a first-order structural transition from monoclinic 1T′ to orthorhombic Td commonly reported in the range of 220–240 K [3,6,13,15,23,25]. Moreover, on cooling, the $\rho_c(T)$ curve shows a clear metal-insulator transition (MIT) around ~20 K, followed by an insulator-metal transition (IMT) below 4 K (inset to Fig. 3(b)). $\rho_c(T)$ in Fig. 3(b) in the 50–200 K range is fitted with a FL $T^2$ term and an e–ph $T^3$ contribution, as shown in the main panel in Fig. 3(b) by a solid red line (labelled FL). $\rho_c(T)$ in the 5–50 K window (where $\frac{d\rho}{dT} < 0$ below 20 K) is captured by incorporating 3D WL corrections ($T^{p/2}$) to the e–e quantum correction ($T^{1/2}$) given by the conductance $G(T)=G_0+mT^{1/2}+BT^{p/2}$ [58,59], where $m$, $B$, $G_0$, and $p$ are fitting parameters [39,58,59]. A fit is shown by a solid blue line denoted by modified WL (MWL) in Fig. 3(b). The fitting yields $p$ ~4, $G_0=0.67\pm3.12\times10^{-4}$ $\Omega^{-1}$, $m=7.34\times10^{-3}\pm1.10\times10^{-4}$ $\Omega^{-1}K^{-1/2}$, and $B=-1.943\times10^{-5}\pm3.10\times10^{-7}$ $\Omega^{-1}K^{-2}$. A dephasing rate scaling as $p$ ~4 is characteristic of the dirty-limit e–ph dephasing within the Sergeev–Mitin vibrating-impurity mechanism [59,60]. Compression along the c-axis reduces the interlayer spacing connected by weak van der Waals bonds and can enhance elastic disorder through stacking-fault formation, strain fields, and pressure-activated point defects, where increased disorder enhances temperature-dependent dephasing associated with e–ph scattering [60-62]. The IMT below 4 K (inset of Fig. 3(b)) can be rationalized in terms of two non-exclusive scenarios. First, prior studies report pressure-enhanced superconducting correlations in the 1T′ and Td phases of MoTe$_2$ near 10 GPa [25], whereas bulk 2H-MoTe$_2$ shows no clear superconductivity up to at least 40 GPa [57]. At the same time, first-principles calculations predict pressure-induced metallization of Mo-based systems without a structural transformation of the 2H lattice, even at much higher pressures [56]. Within this framework, the emergence of a small superconducting volume fraction or enhanced superconducting fluctuations under compression could suppress the low-temperature resistivity, producing a re-entrant metallic-like tail without the onset of a zero-resistance state [13,63]. Second, a crossover from WL to WAL provides an alternative scenario in Te-based dichalcogenides. As the phase-coherence time increases upon cooling and approaches the SO scattering time, the quantum correction to the conductivity changes sign, reducing resistivity at low temperatures via destructive backscattering [64,65]. The anomaly below 4 K is thus consistent with SOC-driven quantum interference on a pressure-stabilized semimetallic background [11–13,15,23].

To discriminate between these scenarios, $\rho_c(T, H)$ in Fig. 3(c) shows that the low-temperature resistivity downturn is progressively suppressed and fully eliminated by ~1.5 T, while the MWL contribution persists up to 5 T. This contrasting response indicates that the low-temperature correction is fragile to time-reversal symmetry breaking, consistent with quantum interference effects. To track the evolution of this behavior, MR(*H*,*T*) is shown in Fig. 3(d). Below 4 K, the MR exhibits a sharp positive cusp near zero field, well described by a 3D WAL (HLN-like) model with a phase-coherence length $L_\phi \approx 65$ nm [48,66], comparable to values reported for 1T′-MoTe$_2$ [11,13,15]. With increasing temperature, dephasing suppresses the WAL correction and restores a positive, nearly parabolic MR dominated by *e*–*e* interactions. At intermediate temperatures (~10 K), a shallow negative MR emerges at low fields, marking the onset of WL competing with the *e*– $e$ background, before being overwhelmed by the $H^2$ orbital response at higher temperatures (see 15 K in Fig. 3(d)).

MoTe$_2$ resides unusually close in free energy to the boundary separating the semiconducting 2H phase from the distorted metallic 1T′ manifold, such that modest perturbations including strain, disorder, or temperature can substantially stabilize mixed-phase electronic behavior [4]. Under 15.6 GPa, the transport response evolves consistently with this proximity: (i) At high temperatures, the resistivity becomes metallic and exhibits a clear anomaly near ~225 K (Fig. 3(b)), at a temperature scale comparable to the 1T′→Td structural transition reported in MoTe$_2$ systems. (ii) At lower temperatures, the resistivity develops an upturn governed by quantum corrections, indicating a pressure-induced MIT-like crossover rather than activated semiconducting transport. (iii) The low-field MR further reveals a crossover from WAL to WL, together with strong $e$–$e$ interaction contributions. These features, taken together, are characteristic of transport in the 1T′/Td manifolds of MoTe$_2$, but not of the pristine 2H semiconducting phase. Within this picture, the pressure-tuned system enters a 1T′/Td-like electronic manifold, as inferred from transport, in which semimetallic behavior and quantum interference coexist with disorder-assisted localization, without requiring a fully resolved structural transition. This reflects a mixed or hybrid electronic regime that emerges from the close energetic proximity of competing structural phases.

First-principles calculations including SOC in Figs. 3(e)-(j) reveal a continuous pressure-driven evolution of the electronic structure in 2H-MoTe$_2$. At ambient pressure, the system exhibits an indirect bandgap (~0.69 eV), consistent with a semiconducting 2H phase with spin-degenerate bands due to preserved inversion and time-reversal symmetries [5,30]. With increasing pressure, the bandgap decreases monotonically and collapses near ~12.5 GPa, giving rise to a semimetallic electronic structure that persists at higher pressures. This smooth evolution, without signatures of an abrupt transition, establishes a continuous pathway from a semiconductor to a semimetallic electronic manifold, consistent with pressure-driven electronic reconstruction predicted for $MoTe_2$ systems [56]. Inclusion of SOC introduces pronounced band splittings and modifies the band curvature near the Fermi level, particularly around the K and M points, yielding a low-energy electronic structure resembling that of semimetallic $MoTe_2$ polymorphs [4,11]. While the calculations do not directly address quantum interference, the emergence of a SOC-influenced semimetallic manifold under pressure provides an electronic structure consistent with the SOC-driven quantum interference effects observed experimentally, as manifested by the WAL behavior.

In summary, hydrostatic pressure drives bulk $2H-MoTe_2$ from a strongly insulating state into a semimetallic transport regime with signatures resembling the 1T′/Td phases. At ambient pressure, the out-of-plane resistivity evolves from metallic behavior to activated transport and 3D Mott variable-range hopping, while a pronounced, non-saturating magnetoresistance near 45 K marks the competition between extended and localized transport channels. In the hopping regime, magnetoresistance is captured by a phenomenological form incorporating quantum interference and wave-function-shrinkage contributions. Under 15.6 GPa, a semimetallic state emerges with quantum corrections, including a SOC-driven WAL cusp. First-principles calculations reveal a continuous pressure-driven bandgap collapse. The comparable interference and coherence length scales establish a unified disorder-controlled transport framework and provide a route to testing consistency across characteristic length scales.

**Acknowledgements**

M.H., M.E., H.S. & S.E.K. acknowledge the support from the American University of Sharjah – University of Sharjah Joint Research Program (JRP25-004). M. H., G. L. and A.A acknowledge the support from the ARG01-0516-230179 project funded by Qatar Research Development and Innovation Council (QRDI).

**Figure Captions**

**FIG. 1.** (a) X-ray diffraction along the *c*-axis showing the (00ℓ) reflections. (b) Raman spectrum (514 nm excitation) with characteristic phonon modes; inset: 2H crystal structure. (c) $\rho_c(T)$ at ambient pressure. Solid lines represent fits to FL, Arrhenius, and 3D Mott VRH regimes; dashed extrapolations intersect near 45 K. Inset: series combination of the FL and VRH. (d) $\rho_c(T)$ and $\rho_{ab}(T)$ at 0 and 9 T (left axis), and corresponding MR(*T*) for *H*//*c* and *H*//*ab* (right axis), showing a maximum near 45 K.

**FIG. 2.** (a) MR(*H*) measured at selected temperatures for *H*//*c* and *H*//*ab* up to 60 T. (b) MR(*T*) at 9 T for *H*//*c* and *H*//*ab*. The dashed vertical line marks the 45 K crossover. (c) $MR_c(H)$ at selected temperatures; inset: ES fit (solid red line) to Eq. (2), and the extracted $\xi$. (d) Low-temperature MR(*H*) with fits to Eq. (1) in solid red lines. (e-g) Temperature evolution of the *P*, *A*, and $L_{\text{loop}}$; solid lines are guides to the eye.

**FIG. 3.** (a) $\rho_c(T)$ under listed hydrostatic pressures. (b) $\rho_c(T)$ at 15.6 GPa, showing FL behavior at high temperatures (red line) and MWL at low temperatures (blue line); inset: low-temperature zoom with MWL fit. Vertical dashed lines mark the crossover near ~4 K. (c) $\rho_c(T)$ at 15.6 GPa measured under the listed applied magnetic fields; solid blue lines denote MWL fits. (d) $MR_c(H)$ at 15.6 GPa in the range ± 5 T at selected temperatures; the 1.9 K curve is fitted to a 3D WAL model (solid red line). (e–j) First-principles electronic band structures including SOC under hydrostatic pressure from 0 to 15 GPa; the energy gap evolution between the conduction-band minimum and valence-band maximum is indicated by arrows in (e-h) panels, and the Fermi level $E_F$ is set to zero in all panels.

**FIG. 1.**

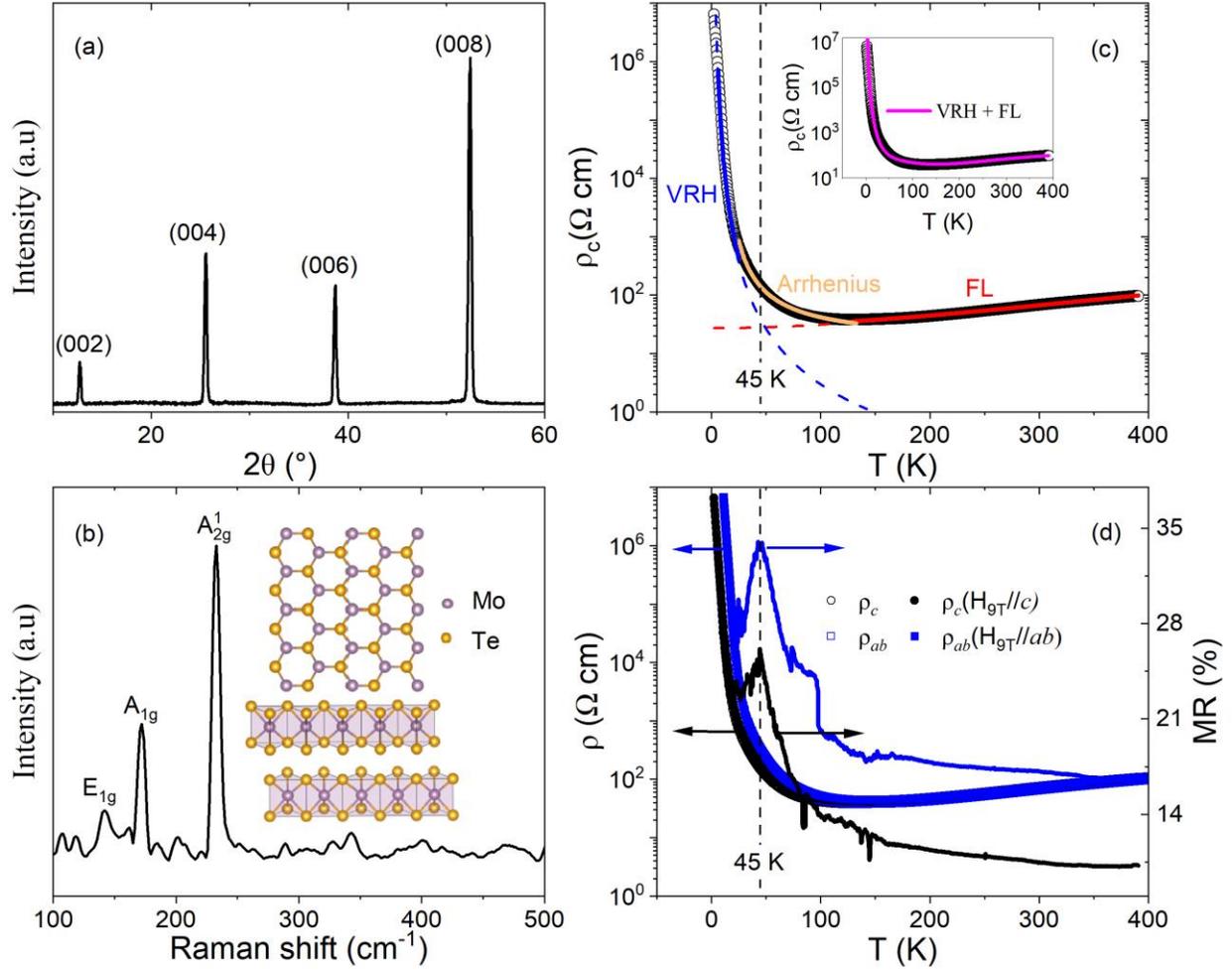

**FIG. 2.**

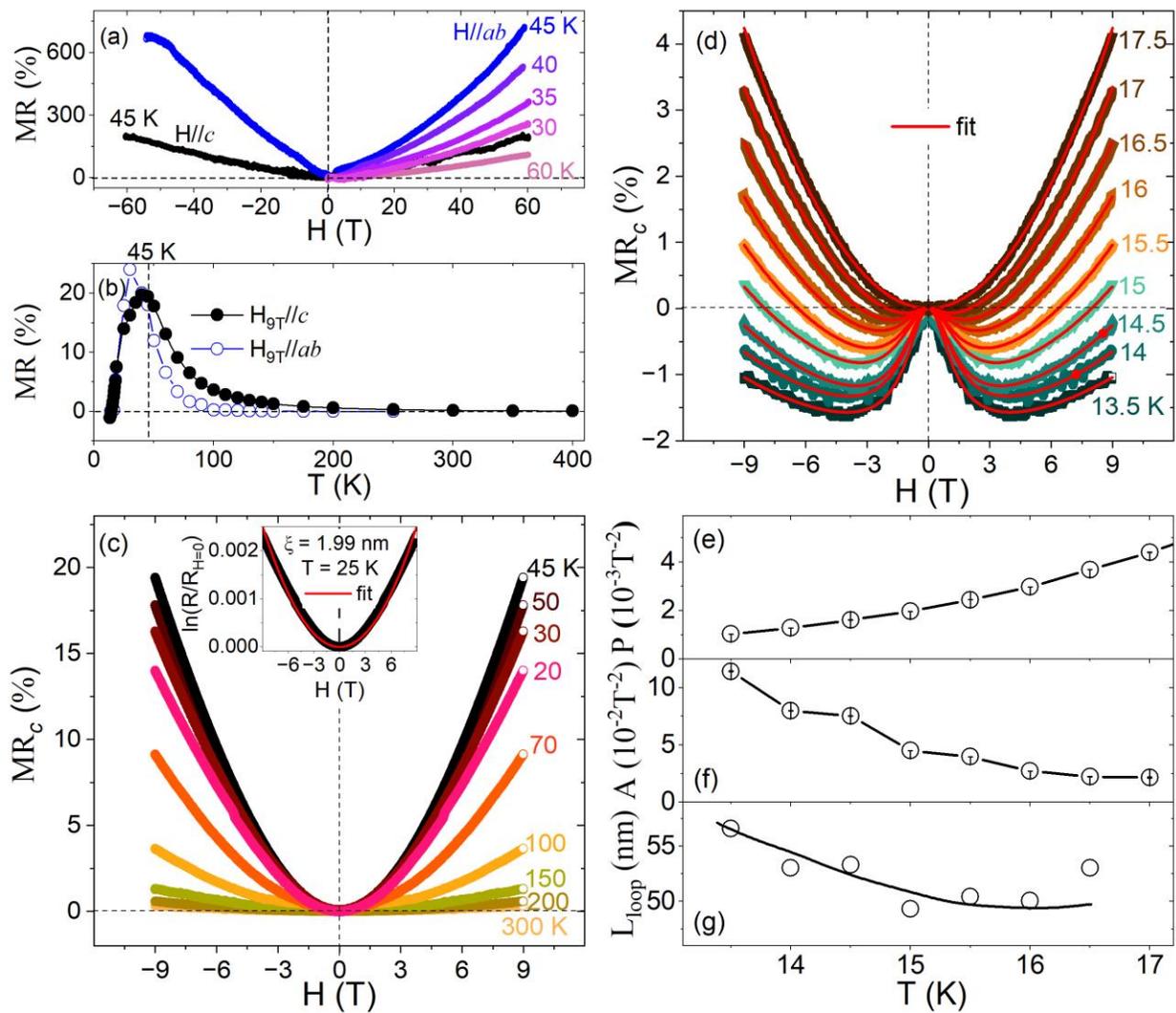

**FIG. 3.**

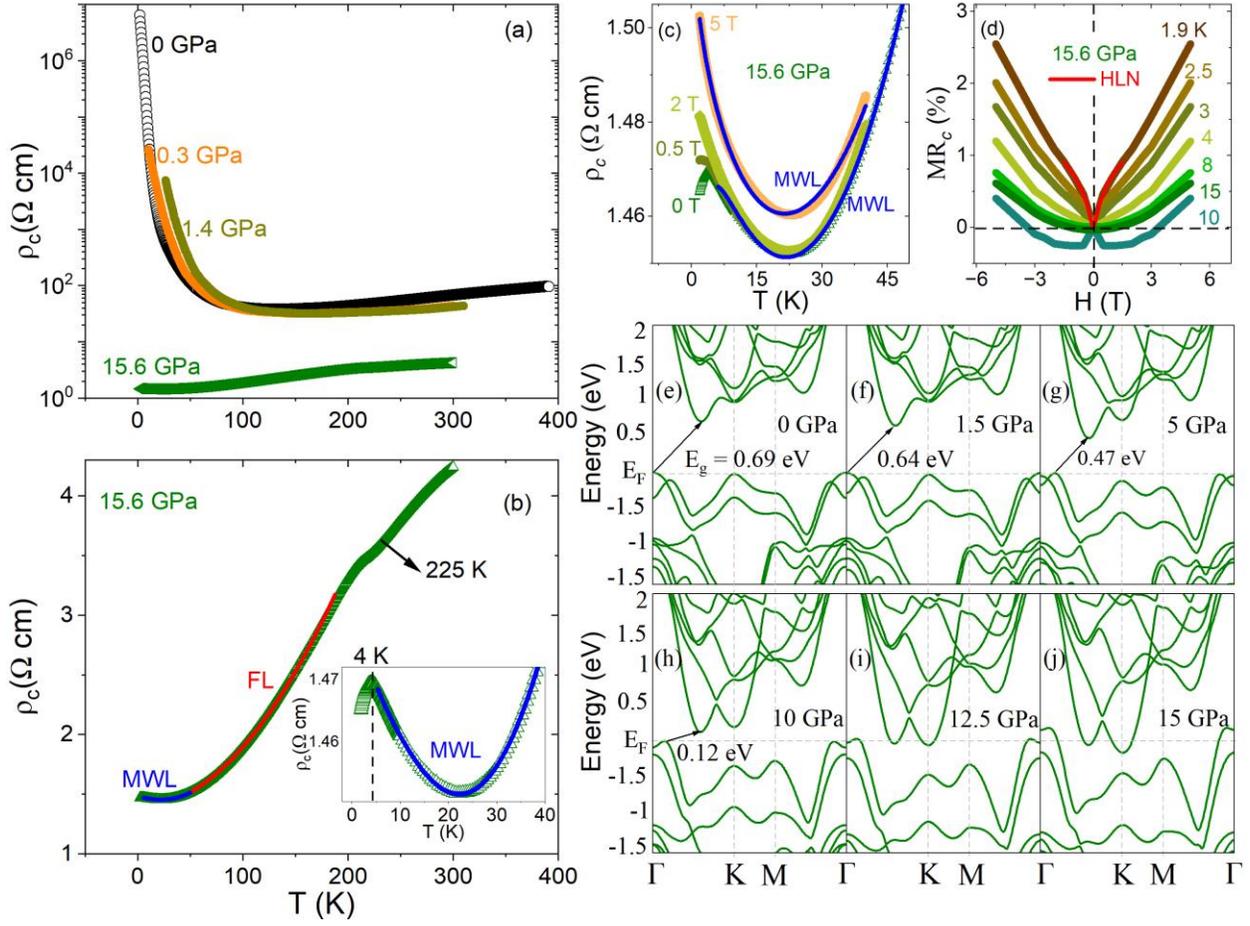